\documentclass[aps,prc,preprintnumbers,twocolumn,superscriptaddress,showpacs,nofootinbib]{revtex4}
\usepackage{graphicx,color,amsmath,amssymb,bm}

\newcommand{\eq}[1]{Eq.~(\ref{#1})}
\newcommand{\be}{\begin{equation}}
\newcommand{\ee}{\end{equation}}
\newcommand{\bea}{\begin{eqnarray}}
\newcommand{\eea}{\end{eqnarray}}
\newcommand{\GV}{G_V}

\begin{document}

\preprint{NT@UW-09-22}

\title{Isospin-symmetry-breaking corrections to superallowed Fermi
$\beta$ decay: \\ 
Radial excitations}

\author{G.\ A.\ Miller}
\email[E-mail:~]{miller@phys.washington.edu}
\affiliation{Department of Physics, University of Washington,
Seattle, WA 98195-1560}
\author{A.\ Schwenk}
\email[E-mail:~]{schwenk@triumf.ca}
\affiliation{TRIUMF, 4004 Wesbrook Mall, Vancouver, BC, V6T 2A3, Canada}


\begin{abstract}
Based on an exact formalism, we study the effects of
isospin-symmetry-breaking interactions on superallowed $0^{+}
\rightarrow 0^{+}$ transitions. We calculate the second-order
renormalization of the Fermi matrix element due to radial
contributions and show that radial excitations neglected in the
treatment of Towner and Hardy are significant. These are 
estimated to decrease the isospin-symmetry-breaking corrections.
Our results provide a correction term that can be included in 
existing approaches.
\end{abstract}


\maketitle

\section{Introduction}

Superallowed nuclear $\beta$ decays play a key role for precision
tests of fundamental symmetries: They provide the most stringent test
of the conserved-vector-current (CVC) hypothesis, the best limit on
scalar interactions, and the most precise value for the up-down
Cabibbo-Kobayashi-Maskawa (CKM) matrix element
$V_{ud}$~\cite{HT09}. The precise extraction of $V_{ud}$ from
superallowed transitions requires the evaluation of theoretical
corrections due to isospin-symmetry breaking (ISB)~\cite{TH08,OB95}
and radiative~\cite{MS06} effects with an uncertainty of $10\%$, to
guarantee a desired accuracy of $0.1\%$ for $V_{ud}$. Because of the
very high precision reached experimentally, the uncertainty of
$V_{ud}$ is currently limited by ISB and radiative
corrections~\cite{HT09}. This presents a challenge for nuclear theory.

Superallowed $0^{+}$~$\rightarrow$~$0^{+}$ Fermi $\beta$ decays depend
only on the vector part of weak interactions, and with CVC, the
transition ``$ft$ value'' should be nucleus-independent:
\be
ft = \frac{2 \pi^3 \hbar^7 \ln 2}{| M_F |^2 \, \GV^2 m_e^5 c^4} =
{\rm const.} \,,
\ee
where $\GV $ is the vector coupling constant and $M_F$ is the Fermi
matrix element. The CVC hypothesis depends on the assumption of isospin
symmetry, which is not exact in nuclei, but broken by electromagnetic
and quark mass effects. As a result, $M_F$ is reduced from the
symmetry value, $M_0 = \sqrt{2}$ for $T=1$ parent and daughter
states, by ISB corrections $\delta_C$,
\be
|M_F|^2 = |M_0|^2 \, ( 1 - \delta_C ) \,.
\ee
In addition, there are radiative corrections, but we focus on
$\delta_C$ in this paper.

Towner and Hardy (TH) have shown~\cite{TH08,HT09} that the calculated
ISB corrections eliminate much of the considerable scatter present in
the uncorrected $ft$ values, and the statistical consistency among the
corrected ${\mathcal F}t$ values 
is some evidence that the calculated ISB corrections are
not unreasonable. However, the precise extraction of $V_{ud}$ and the
importance of testing the Standard Model have stimulated us to
undertake a reevaluation based on an exact formalism~\cite{MS08}.
In addition, our goal is to connect the calculated corrections to the
accurate understanding of ISB in nuclear forces~\cite{ISB1,ISB2}.
Reference~\cite{MS08} showed that there are specific corrections to
the treatment of TH~\cite{TH08} and explained these using schematic
models without making numerical estimates. In particular, radial
excitations are neglected by TH, and as a result the transition
operator violates the isospin commuation relations. The TH approach
is discussed in more detail in Sect.~\ref{THapproach}.

Following our work, there have been a number of theoretical
developments: Auerbach studied ISB corrections assuming that the
dominant physics is due to the isovector monopole
resonance~\cite{Au09}. Assuming that certain reduced matrix elements
of the isosvector part of the Coulomb potential are identical,
Auerbach showed that ISB corrections vanish unless one takes into
account the energy differences between components of the isovector
monopole state of different isospin. Using schematic models, the
resulting estimates for $\delta_C$ are considerably smaller than the
TH results. Liang {\it et al.}~carried out random-phase-approximation
calculations based on relativistic density functionals. They also
obtain smaller values for $\delta_C$ and their corrected ${\mathcal
F}t$ values are statistically consistent~\cite{Liang}.  In addition,
Satula {\it et al.}~have analyzed isospin mixing and implemented
isospin projection in density-functional calculations~\cite{Satula}.

At the same time, the experimental precision has been improved in
several cases~\cite{HT09} and a recent branching ratio measurement for
$^{32}$Ar~\cite{Bhatta} has improved the $ft$ value for this $T=2$,
$T_z=-2$ decay to $0.8 \%$, a precision that is nearing the $0.3 \%$
standard for inclusion into the set of well-known $T=1$ decays. This
is especially interesting, because ISB corrections appear to be larger
for $T=2$ superallowed transitions.

In this work, we start from the exact formalism developed in
Ref.~\cite{MS08}, which is reviewed in Sect.~\ref{exact}. In
Sect.~\ref{relate} we write the correct isospin operator as a sum of
the TH operator and a correction term due to radial excitations. We
show that the neglected radial contributions are significant at
leading (second) order in ISB interactions. In addition, for certain
conditions the correction term cancels ISB corrections to Fermi matrix
elements. This is similar to the case of Auerbach when the components
of the isovector monopole state are degenerate~\cite{Au09}. Our
results demonstrate explicitly the importance of radial excitations
and we discuss the implications in Sect.~\ref{implications}.

\section{TH approach to ISB corrections}
\label{THapproach}

With CVC, the matrix elements of weak vector interactions in nuclei
are not modified by nuclear forces, except for corrections due to ISB
effects. Therefore, one has to evaluate the contributions from
electromagnetic and charge-dependent strong interactions to the Fermi
matrix element $M_F = \langle f | \tau_+ | i \rangle$ between the
initial and final states for superallowed $\beta$ decay, $| i \rangle$
and $| f \rangle$, respectively. Here $\tau_+$ denotes the isospin
raising operator.

Towner and Hardy~\cite{TH08} use a second-quantized formulation
to write the Fermi matrix element as
\be
M_F = \sum_{\alpha, \beta} \langle f | a_{\alpha}^{\dag} a_{\beta}
| i \rangle \langle \alpha | \tau_+ | \beta \rangle \,,
\label{MFq}
\ee
where $a_{\alpha}^{\dag}$ creates a neutron in state $\alpha$ and
$a_{\beta}$ annihilates a proton in state $\beta$. Thus, the label
$\alpha$ is used to denote neutron creation and annihilation
operators, while $\beta$ is used for those of the proton. This
notation is different from the standard notation~\cite{DW}, in which
$b_\alpha$ is used to denote proton annihilation operators.  The
single-particle matrix element $\langle \alpha | \tau_+ | \beta
\rangle$ is assumed to be given by
\be
\langle \alpha | \tau_+ | \beta \rangle
= \delta_{\alpha, \beta} \int_0^{\infty}
R_{\alpha}^n(r) \, R_{\beta}^p(r) \, r^2 dr
\equiv \delta_{\alpha, \beta} \, r_{\alpha} \,,
\label{radi}
\ee
where $R_{\alpha}^n(r)$ and $R_{\beta}^p(r)$ are the neutron and
proton radial wave functions, respectively. Because the radial quantum
numbers of the states $\alpha$ and $\beta$ are set to be the same,
Eq.~(\ref{radi}) assumes that $\tau_+$ creates a neutron in the
single-particle state with the same quantum numbers as those of the
annihilated proton. This is not the case in the presence of ISB. As a
result, the operator in \eq{MFq} is not the correct isospin operator
and the Standard Model isospin commutation relations are lost.

We observe that Eqs.~(\ref{MFq}) and~(\ref{radi}) correspond to the
second-quantized isospin operators
\bea
\tau_+ &=& \sum_{\alpha, \beta} \delta_{\alpha, \beta} \, r_\alpha \,
a^\dagger_\alpha a_\beta \,, \\[1mm]
\tau_- &=& \tau_+^\dagger = \sum_{\alpha, \beta} \delta_{\alpha,
\beta} \, r^*_{\alpha} \, a^\dagger_{\beta} a_{\alpha} \,,
\eea
so that the commutation relations are given by
\be
[ \tau_+ , \tau_- ] = \sum_\alpha |r_\alpha|^2 \, a^\dagger_\alpha
a_\alpha - \sum_\beta |r_\beta|^2 \, a^\dagger_\beta a_\beta \ne
\tau_0 \,,
\ee
which shows explicitly that the Standard Model isospin commutation
relations are violated if one uses the isospin operator of TH. The
violation is due to neglecting parts of the proton wave functions that
are in radial excitations when expanded in a neutron basis. This formal
problem motivated us to develop an exact formalism for ISB corrections
and to study corrections to the TH treatment~\cite{MS08}.

For the calculation of $\delta_C$, TH~\cite{TH08} proceed by
introducing into Eq.~(\ref{MFq}) a complete set of states for the
$(A-1)$-particle system, $|\pi \rangle $, which leads to
\be
M_F = \sum_{\alpha, \pi} \langle f | a_{\alpha}^{\dag} | \pi \rangle
\langle \pi | a_{\alpha} | i \rangle \, r_{\alpha}^{\pi} \,.
\ee
The TH model thus allows for a dependence of the radial integrals on
the intermediate state $\pi$.  If isospin were an exact symmetry, the
matrix elements of the creation and annihilation operators would be
related by hermiticity, $\langle \pi | a_{\alpha} | i \rangle =
\langle f | a_{\alpha}^{\dag} | \pi \rangle^*$, and all radial
integrals would be unity. Hence the symmetry-limit matrix element in
the TH model is given by
\be
M_0 = \sum_{\alpha, \pi} | \langle f | a_{\alpha}^{\dag} | \pi \rangle |^2 \,.
\ee

Towner and Hardy divide the contributions from ISB into two
terms. First, the hermiticity of the matrix elements of $a_{\alpha}$
and $a_{\alpha}^{\dag}$ will be broken, and second, the radial
integrals will differ from unity.  Assuming both effects are small, TH
separate the resulting ISB corrections into two model-dependent
parts~\cite{TH08}
\be
\delta_C = \delta_{C1} + \delta_{C2} \,,
\label{dc1and2}
\ee
where in evaluating $\delta_{C1}$ all radial integrals are set to
unity but the matrix elements are not assumed to be related by
hermiticity, and in evaluating $\delta_{C2}$ it is assumed that
$\langle \pi | a_{\alpha} | i \rangle = \langle f | a_{\alpha}^{\dag}
| \pi \rangle^*$ but $r_{\alpha}^{\pi} \ne 1$.

While \eq{MFq} accounts for important effects of the Coulomb
interaction on the radial wave functions, radial excitations are
neglected in \eq{radi}. We will show in Sect.~\ref{relate} that radial
excitations contribute at the same (second) order in ISB interactions
and are estimated to decrease the ISB corrections $\delta_{C2}$
obtained using the TH isospin operator.

\section{Exact formalism and theorems for ISB corrections}
\label{exact}

In this section, we recall aspects of the formalism and theorems of
Ref.~\cite{MS08}. These show that there are no first-order ISB
corrections to Fermi matrix elements and provide a basis for a
complete second-order calculation. The formalism starts from the
correct isospin operator
\be
\tau_+ = \sum_\alpha a_\alpha^\dagger b_\alpha \,,
\label{true}
\ee
where $\alpha$ represents any single-particle basis, and
$a^\dagger_\alpha$ creates neutrons and $b_\alpha$ annihilates protons
in state $\alpha$. The Fermi matrix element is then given by
\be
M_F = \langle f | \tau_+ | i \rangle \,,
\label{truemf}
\ee
where $|i\rangle$ and $|f\rangle$ are the exact initial and final
eigenstates of the full Hamiltonian $H=H_0+V_C$, with energy $E_i$ and
$E_f$, respectively. Here $V_C$ denotes the sum of all
interactions that do not commute with the vector isospin operator
${\bf T} = \sum_{i=1}^A {\bm \tau}_i/2$,
\be
[H , {\bf T}] = [V_C , {\bf T}] \ne 0
\quad \text{and} \quad [H_0 , {\bf T}] = 0 \,.
\ee
We use round states, $|i)$ and $|f)$, to denote the bare initial and
final eigenstates of $H_0$. Obtaining these states requires a solution
of the $A$-body problem.

The full initial and final eigenstate $|i\rangle$ and $|f\rangle$
can then be written as
\bea
| i \rangle &=& \sqrt{Z_i} \, \biggl[ |i) + \frac{1}{E_i-\Lambda_i H
\Lambda_i}\, \Lambda_i V_C \, |i) \biggr] \,, \label{initial} \\[1mm]
| f \rangle &=& \sqrt{Z_f} \, \biggl[ |f) + \frac{1}{E_f-\Lambda_f H
\Lambda_f}\, \Lambda_f V_C \, |f) \biggr] \,, \label{final}
\eea
with projectors $\Lambda_i \equiv 1-|i)(i|$ and $\Lambda_f =
1-|f)(f|$. The unperturbed states $|i)$ and $|f)$ are related by an
isospin rotation, so that $|f)$ is the isobaric analog state of $|i)$:
\be
|f) = \frac{1}{\sqrt{2T}} \, \tau_+ \, |i)
\quad \text{and} \quad
|i) = \frac{1}{\sqrt{2T}} \, \tau_- \, |f) \,,
\label{analog}
\ee
where $T$ denotes the isospin of the states $|i)$ and $|f)$, and for
simplicity we consider parent cases with isospin projection
$T_z=(N-Z)/2=-T$ or $T=1$, $T_z=0$, so that $M_0 = ( f | \tau_+ | i )
= \sqrt{2T}$. The factors $Z_i$ and $Z_f$ are taken to be real and
ensure that the full eigenstates are normalized. With $(f| \, \tau_+
\Lambda_i = 0$ and $\Lambda_f \tau_+ \, |i)=0$, the exact Fermi matrix
element is given by
\begin{multline}
M_F = \sqrt{Z_i Z_f} \, \biggl[ M_0 + (f| \, V_C \Lambda_f \,
\frac{1}{E_f-\Lambda_f H \Lambda_f} \\[1mm]
\times \tau_+ \, \frac{1}{E_i-\Lambda_i H \Lambda_i} \, \Lambda_i V_C
\, |i) \, \biggr] \,.
\label{th1}
\end{multline}
This is the first theorem~\cite{MS08} and, because of $Z_{i,f} = 1 +
{\mathcal O} (V_C^2)$, demonstrates that there are no first-order ISB
corrections to $M_F$.

Another formulation is obtained by expanding in the difference of the
charge-dependent interactions $\Delta V_C$ between the initial
proton-rich and final neutron-rich states. In this case, the full
Hamiltonian is given by
\be
H = \widetilde{H}_0 + \Delta V_C \,,
\ee
where $\widetilde{H}_0$ includes the effects of $V_C$ common to the
initial and final states, for example the Coulomb interactions in the
core, and $\Delta V_C$ takes into account all charge-dependent
interactions of the extra proton with the other nucleons in the
initial state. In this formulation, the bare states, $|i)$ and $|f)$,
are eigenstates of $\widetilde{H}_0$, but are not eigenstates of the
isospin operator:
\bea
|i) &=& \sum_{T' \geqslant |T_z|} \gamma_{T'} \, |T',T_z) \,, 
\label{analog1} \\
|f) &=& \sum_{T' \geqslant |T_z+1|} \gamma_{T'} \, |T',T_z+1) \,.
\label{analog2}
\eea
In the isospin-symmetry limit, $\gamma_{T'}=\delta_{T',T}$, where $T$
is the isospin of the bare states of \eq{analog} (which is
also the dominant isospin in the presence of ISB).

In this case, the full eigenstates can be written as
\be
|f\rangle = |f) \quad \text{and} \quad
|i\rangle =
\sqrt{Z} \, |i) + \frac{1}{E_i-\Lambda_i \widetilde{H}_0 \Lambda_i} \,
\Lambda_i \Delta V_C \, |i\rangle \,,
\label{states}
\ee
and we obtain for the exact Fermi matrix element
\begin{multline}
M_F = \sqrt{Z} \: \sum_{T'} \: |\gamma_{T'}|^2
\sqrt{T'(T'+1)-T_z(T_z+1)} \\
+ (f| \tau_+ \Lambda_i \, \frac{1}{E_i - \Lambda_i \widetilde{H}_0
\Lambda_i} \, \Lambda_i \, \Delta V_c \, |i\rangle \,, \label{th2}
\end{multline}
where the sum is over $T' \geqslant \max(|T_z|,|T_z+1|)$.
The $|T, T_z)$ expansion of the states $|i)$ and $|f)$ presents a more
careful evaluation of the second theorem of Ref.~\cite{MS08}. Since
$\gamma_T = 1 + {\mathcal O}(V_C^2)$ and $\gamma_{T' \neq T} =
{\mathcal O}(V_C)$, it follows that $(f| \tau_+ \Lambda_i$ is of
first order in ISB interactions. Combined with $Z = 1 + {\mathcal O}
\bigl((\Delta V_C)^2\bigr)$, \eq{th2} explicitly shows that ISB
corrections to $M_F$ start at second order. In the following, we
will work with the first formulation.

\section{Relation between TH operator and isospin}
\label{relate}

Next we derive a relation between the TH operator and the correct
isospin operator based on the exact formalism. We use basis states
given by conveniently-chosen one-body potentials $U$ and $U+U_C$,
where $U_C$ accounts for charge-dependent effects. The single-particle
(sp) potentials are chosen to minimize the effects of residual
interactions,
\be
\Delta V = V + V_C - (U + U_C) \,,
\ee
so that the Hamiltonian is given by
\be
H = H_{\rm sp} + \Delta V \quad {\rm with} \quad
H_{\rm sp} = T + U + U_C \,.
\label{Hsp}
\ee

We express the isospin raising operator $\tau_+$ in a mixed
representation, where $|\alpha\rangle$ denotes the eigenstates of
the single-particle Hamiltonian $H_{\rm sp}$ and $|\widetilde{\alpha}
\rangle$ the eigenstates of the isospin-symmetric part $T + U$.
The creation operators in the two bases are related by
\be
a^\dagger_{\alpha} = \sum_{\alpha'} \,
a^\dagger_{\widetilde{\alpha'}} \,
\langle\widetilde{\alpha'}|\alpha\rangle \,,
\label{oprel}
\ee
where the tilde indicates the basis and the sum is over all
single-particle quantum numbers. The correct isospin operator,
\eq{true}, can then be expressed as
\be
\tau_+ = \sum_{\alpha,\alpha'} \,
a^\dagger_{\widetilde{\alpha'}} \,
\langle\widetilde{\alpha'}|\alpha\rangle \,
b_\alpha \,, \label{mixed}
\ee
which TH use as their starting point. However, \eq{mixed} allows the
states $|\alpha\rangle$ and $|\widetilde{\alpha'}\rangle$ to have
different radial quantum numbers $n$ and $n'$, to be explicit
\be
\alpha = nljm \quad {\rm and} \quad \alpha' = n'ljm \,,
\ee
with orbital angular momentum $l$, total angular momentum $j=l \pm
1/2$, and magnetic quantum number $m$.

The TH operator $\tau_+^{\rm TH}$ of Eqs.~(\ref{MFq}) and~(\ref{radi})
is obtained by keeping the terms with $\alpha=\alpha'$,
\be
\tau_+^{\rm TH} = \sum_{\alpha} a^\dagger_{\widetilde{\alpha}} \,
b_{\alpha} \, r_\alpha \,,
\ee
with $r_\alpha = \langle \widetilde{\alpha}|\alpha\rangle$ in the TH
notation. Therefore, we define the correction operator,
\be
\delta\tau_+ = \sum_{\alpha,\alpha' \neq \alpha}
a^\dagger_{\widetilde{\alpha'}} \, b_{\alpha} \, \langle
\widetilde{\alpha'}|\alpha\rangle \,.
\label{dtau}
\ee
Then the correct isospin operator and the exact Fermi matrix element
are given by
\bea
\tau_+ &=& \tau_+^{\rm TH} + \delta\tau_+ \,, \\[1mm]
M_F &=& \langle f|\tau_+^{\rm TH}|i\rangle + 
\langle f|\delta\tau_+|i\rangle \,. \label{twot}
\eea

We evaluate both terms in \eq{twot} to second order in ISB
interactions. This will explicitly demonstrate that the second
term due to radial excitations is of the same (second) order as
the TH term. We start with the latter,
\be
M_F^{\rm TH} = \langle f|\tau_+^{\rm TH}|i\rangle 
= M_0 - \langle f| \sum_{\alpha} a_{\widetilde{\alpha}}^\dagger \,
b_\alpha \, (1-r_\alpha) |i\rangle \,.
\label{MFTH}
\ee
Here we have applied the TH procedure and neglected the isospin-mixing
correction $\delta_{C1}$ of \eq{dc1and2}, which is small in
Ref.~\cite{TH08}, so that $M_F^{\rm TH}=M_0$ for $r_\alpha=1$. Since
$(1-r_\alpha)$ starts at second order in ISB interactions, we can
replace the full eigenstates in the second term in \eq{MFTH} by $|i)$
and $|f)$. Using a single-particle version of \eq{states}, we express
$(1-r_\alpha)$ in terms of the matrix elements of the one-body
potential $U_C$, that accounts for the difference between the
$|\widetilde{\alpha}\rangle$ and $|\alpha\rangle$ basis states, to
second order,
\be
1 - r_\alpha \approx \frac{1}{2} \sum_{\alpha' \neq \alpha}
\frac{\bigl|\langle \widetilde{\alpha'}|U_C|\widetilde{\alpha}
\rangle\bigr|^2}{\bigl(\widetilde{E}_\alpha-\widetilde{E}_{\alpha'}
\bigr)^2} \,,
\ee
where $\widetilde{E}_\alpha$ are the eigenvalues of the
isospin-symmetric single-particle Hamiltonian $T+U$. For simplicity,
we take $|i)$ to be a $Z-N$ proton plus
core configuration. We define the occupation probabilities
$\widetilde{\rho}_\alpha$ of the proton excess in the
$|\widetilde{\alpha}\rangle$ basis, normalized so that $\sum_\alpha
\widetilde{\rho}_\alpha=(i|\tau_- \tau_+ |i)=2T$.\footnote{To clarify
the notation, the symmetry-limit matrix element can then be written as
\bea
M_0 &=& \frac{1}{\sqrt{2 T}} \, (i| \, \sum_\alpha
b^\dagger_{\widetilde{\alpha}} \, a_{\widetilde{\alpha}} \,
\sum_\beta a^\dagger_{\widetilde{\beta}} \, b_{\widetilde{\beta}}
\, |i) \,, \\
&=& \frac{1}{\sqrt{2 T}} \, (i| \, \sum_\alpha
b^\dagger_{\widetilde{\alpha}} \, b_{\widetilde{\alpha}} \, |i)
= \frac{1}{\sqrt{2 T}} \sum_\alpha \widetilde{\rho}_\alpha \,,
\eea
where the $|\widetilde{\alpha}\rangle$ basis, appropriate for the
state $|i)$, was used for $\tau_+$ and $\tau_-$.} As a
result, we find for the TH term
\be
M_F^{\rm TH} \approx M_0 - \frac{1}{2} 
\frac{1}{\sqrt{2T}} \sum_{\alpha,\alpha'\neq\alpha}
\widetilde{\rho}_\alpha \, 
\frac{\bigl|\langle \widetilde{\alpha'}|U_C|\widetilde{\alpha}
\rangle\bigr|^2}{\bigl(\widetilde{E}_\alpha-\widetilde{E}_{\alpha'}
\bigr)^2} \,.
\label{tha}
\ee
In the limit of sharp occupation probabilities
$\widetilde{\rho}_\alpha$, the $\alpha$ ($\alpha'$) sum is over
occupied (unoccupied) states.

Next we evaluate the contributions due to radial excitations,
$\delta M_F = \langle f|\delta\tau_+|i\rangle$. To second order
in ISB interactions, we have
\bea
\delta M_F &\approx& (f|\delta \tau_+|i) + \frac{1}{\sqrt{2T}} \, ( i
| \, \tau_- \, \delta \tau_+ \frac{1}{E_i - \Lambda_i H \Lambda_i} \,
\Lambda_i V_C |i ) \nonumber \\[1mm]
&+& \frac{1}{\sqrt{2T}} \, ( i | \, \tau_- \, V_C \Lambda_f \,
\frac{1}{E_f-\Lambda_f H \Lambda_f} \, \delta \tau_+ \, | i ) \,.
\label{dMF}
\eea
While we will estimate $\delta M_F$ making similar approximations
as for $M_F^{\rm TH}$ of \eq{tha}, the result \eq{dMF} provides
a correction term that can be included in future numerical
calculations
of ISB corrections. For $\delta \tau_+$ we also need the overlap
$\langle\widetilde{\alpha'}|\alpha\rangle$ for $\alpha' \neq \alpha$,
which starts at first order,
\be
\langle\widetilde{\alpha'}|\alpha\rangle \approx 
\frac{1}{\widetilde{E}_\alpha
-\widetilde{E}_{\alpha'}} \, \langle\widetilde{\alpha'}|
U_C |\widetilde{\alpha}\rangle \,.
\ee

We start with the first term of \eq{dMF},
\bea
(f|\delta \tau_+|i) &=& \frac{1}{\sqrt{2 T}} \,
(i| \, \sum_\beta b^\dagger_{\widetilde{\beta}} \,
a_{\widetilde{\beta}} \sum_{\alpha,\alpha' \neq \alpha}
a^\dagger_{\widetilde{\alpha'}} \, b_\alpha \, \langle
\widetilde{\alpha'} | \alpha \rangle \, |i) \,, \nonumber \\
&=& \frac{1}{\sqrt{2 T}} \, (i| \sum_{\alpha,\alpha' \neq \alpha}
b^\dagger_{\widetilde{\alpha'}} \, b_\alpha \, \langle
\widetilde{\alpha'} | \alpha \rangle \, |i) \,,
\eea
where we have used that, for the considered configuration of $|i)$,
the neutron annihilation and creation operators evaluate to
$\delta_{\widetilde{\alpha},\widetilde{\beta}}$. After transforming
$b_\alpha$ to the $|\widetilde{\alpha}\rangle$ basis, using
the Hermitian conjugate of \eq{oprel}, we obtain
\bea
(f|\delta \tau_+|i) &=&
\frac{1}{\sqrt{2 T}} \, (i| \sum_{\alpha,\alpha' \neq \alpha}
b^\dagger_{\widetilde{\alpha'}} \, \sum_\beta b_{\widetilde{\beta}}
\, \langle \alpha | \widetilde{\beta} \rangle \,
\langle \widetilde{\alpha'} | \alpha \rangle \, |i) \,, \nonumber \\
&\approx& \frac{1}{\sqrt{2T}} \sum_{\alpha,\alpha' \neq \alpha}
\widetilde{\rho}_{\alpha'} \, 
\frac{\bigl|\langle \widetilde{\alpha'}|U_C|\widetilde{\alpha}
\rangle\bigr|^2}{\bigl(\widetilde{E}_\alpha-\widetilde{E}_{\alpha'}
\bigr)^2} \,.
\label{dMF1}
\eea

We estimate the second and third terms of \eq{dMF} using a closure
approximation, that is we replace $E_i - \Lambda_i H \Lambda_i$ by
$\Delta E_i<0$, and similarly for $E_f - \Lambda_f H \Lambda_f$. In
addition, we approximate $V_C$ by the ISB one-body potential
$\sum_{\gamma,\gamma' \neq \gamma} \langle \gamma | U_C | \gamma'
\rangle \, b^\dagger_\gamma \, b_{\gamma'}$, where $\gamma \neq
\gamma'$ ensures the action of the projectors $\Lambda_i$ and
$\Lambda_f$. After contracting the neutron operators, we find
for the second term
\bea
&& \frac{1}{\sqrt{2T}} \, ( i | \, \tau_-
\, \delta \tau_+ \frac{1}{E_i - \Lambda_i H \Lambda_i} \,
\Lambda_i V_C \, |i ) \nonumber \\[1mm]
&=& \frac{1}{\sqrt{2 T}} \, (i| \sum_{\alpha,\alpha' \neq \alpha}
b^\dagger_{\widetilde{\alpha'}} b_\alpha \, \frac{
\langle \widetilde{\alpha'} | \alpha \rangle}{\Delta E_i}
\sum_{\gamma,\gamma' \neq \gamma}
\langle \gamma | U_C | \gamma' \rangle \, b^\dagger_\gamma
b_{\gamma'} |i) \,, \nonumber \\
&=& - \frac{1}{\sqrt{2T}} \sum_{\alpha,\alpha' \neq \alpha}
\widetilde{\rho}_{\alpha'} \, 
\frac{\bigl|\langle \widetilde{\alpha'}|U_C|\widetilde{\alpha}
\rangle\bigr|^2}{\bigl|(\widetilde{E}_\alpha-\widetilde{E}_{\alpha'})
\Delta E_i \bigr|} \,.
\label{dMF2}
\eea
For the third term, we obtain
\bea
&& \frac{1}{\sqrt{2T}} \, ( i | \, \tau_- \, V_C \Lambda_f \,
\frac{1}{E_f-\Lambda_f H \Lambda_f} \, \delta \tau_+ \, | i )
\nonumber \\
&=& \frac{1}{\sqrt{2 T}} \, (i| \sum_{\alpha,\alpha' \neq \alpha}
b^\dagger_{\widetilde{\alpha'}} \sum_{\gamma,\gamma' \neq \gamma}
\langle \gamma | U_C | \gamma' \rangle \, b^\dagger_\gamma \,
b_{\gamma'} b_\alpha \, \frac{
\langle \widetilde{\alpha'} | \alpha \rangle}{\Delta E_f} |i) \,.
\nonumber
\eea
For nonzero overlap, this requires two-particle--two-hole
configurations in $|i)$, while the first and second terms receive
contributions at the level of the best Slater determinant. Assuming
residual interactions are weak, we neglect the third term. Combining
Eqs.~(\ref{dMF1}) and~(\ref{dMF2}), we have for the correction term
\bea
\delta M_F &=& \frac{1}{\sqrt{2T}} \sum_{\alpha,\alpha' \neq \alpha}
\widetilde{\rho}_{\alpha'} \, 
\frac{\bigl|\langle \widetilde{\alpha'}|U_C|\widetilde{\alpha}
\rangle\bigr|^2}{\bigl(\widetilde{E}_\alpha-\widetilde{E}_{\alpha'}
\bigr)^2} \nonumber \\
&-& \frac{1}{\sqrt{2T}} \sum_{\alpha,\alpha' \neq \alpha}
\widetilde{\rho}_{\alpha'} \, 
\frac{\bigl|\langle \widetilde{\alpha'}|U_C|\widetilde{\alpha}
\rangle\bigr|^2}{\bigl|(\widetilde{E}_\alpha-\widetilde{E}_{\alpha'})
\Delta E_i \bigr|} \,.
\label{dMFest}
\eea
Comparing our estimate $\delta M_F$ to the corresponding TH term,
\eq{tha}, demonstrates that radial excitations are significant.
The same estimate, \eq{dMFest}, is found when $V_C$ is
approximated by the isovector part of the ISB one-body potential,
$\sum_{\gamma,\gamma' \neq \gamma} \langle \gamma | U_C | \gamma'
\rangle \, (b^\dagger_\gamma \, b_{\gamma'} - a^\dagger_\gamma \,
a_{\gamma'})/2$. In this case, the second term of $\delta M_F$
is $1/2$ of \eq{dMF2}, but the third term also yields $1/2$ of
this (for $\Delta E_i = \Delta E_f$). 

Assuming radial excitations are dominated by $n$ to $n+1$, we have
$\widetilde{E}_\alpha-\widetilde{E}_{\alpha'} = 2 \hbar \omega$, where
$\omega$ is a typical oscillator frequency. Moreover, if the relevant
excitations are dominated by the isovector monopole state, the value
of $|\Delta E_i|$ ranges between $3$ and $4 \hbar
\omega$~\cite{Au09}. For $\widetilde{E}_\alpha-\widetilde{E}_{\alpha'}
= 2 \hbar \omega$ and $|\Delta E_i| = 4 \hbar \omega$, we find that
the correction term completely cancels the TH contribution
$\delta_{C2}$ at second order. This result is similar to the
energy-degenerate case of Auerbach~\cite{Au09}. Our estimate shows
that, if the contributions of the isovector monopole state dominate
ISB, the radial excitations neglected by TH decrease ISB corrections.

\section{Implications for ISB corrections}
\label{implications}

We have used the exact formalism of Ref.~\cite{MS08} to calculate the
renormalization of the Fermi matrix element due to radial
contributions. We expressed the correct isospin operator as a sum of
the TH operator and a correction term involving radial excitations,
which were shown to be significant and estimated to decrease the
ISB corrections $\delta_{C2}$ of Ref.~\cite{TH08}. In addition, for
certain conditions the ISB corrections due to radial contributions can
cancel at second order in ISB interactions. A reduction due to the
correction term implies that the extracted value of $V_{ud}$ may be
reduced. Moreover, our results can provide a possible explanation for
the smaller ISB corrections found in Refs.~\cite{Au09,Liang}, although
these calculations are more exploratory at this stage.

An important direction for future research is to include the
correction term of Eqs.~(\ref{dtau}) and~(\ref{dMF}) in numerical
calculations of ISB corrections that follow the TH approach or
make a similar truncation of basis states.

\acknowledgments

We thank G.\ F.\ Grinyer, J.\ C.\ Hardy and I.\ S.\ Towner for useful
discussions. This work was supported in part by the US Department of
Energy under Grant No.~DE--FG02--97ER41014 and by the Natural Sciences
and Engineering Research Council of Canada (NSERC). TRIUMF receives
funding via a contribution through the National Research Council
Canada.


\begin{thebibliography}{99}
\bibitem{HT09} J.\ C.\ Hardy and I.\ S.\ Towner, \prc {\bf 79},
055502 (2009).

\bibitem{TH08} I.\ S.\ Towner and J.\ C.\ Hardy, \prc {\bf 77},
025501 (2008).

\bibitem{OB95} W.\ E.\ Ormand and B.\ A.\ Brown, \prc {\bf 52},
2455 (1995).

\bibitem{MS06} W.\ J. Marciano and A.\ Sirlin, \prl {\bf 96},
032002 (2006).

\bibitem{MS08} G.\ A.\ Miller and A.\ Schwenk, Phys.\ Rev.\ C
{\bf 78}, 035501 (2008).

\bibitem{ISB1} G.\ A.\ Miller, B.\ M.\ K.\ Nefkens and I.\ Slaus,
Phys.\ Rept.\ {\bf 194}, 1 (1990).

\bibitem{ISB2} G.\ A.\ Miller, A.\ K.\ Opper and E.\ J.\ Stephenson,
Annu.\ Rev.\ Nucl.\ Part.\ Sci.\ {\bf 56}, 253 (2006).

\bibitem{Au09} N.\ Auerbach, Phys.\ Rev.\ C {\bf 79}, 035502 (2009).

\bibitem{Liang} H.\ Liang, N.\ Van Giai and J.\ Meng, \prc {\bf 79},
064316 (2009).

\bibitem{Satula} W.\ Satula, J.\ Dobaczewski, W.\ Nazarewicz and
M.\ Rafalski, \prl {\bf 103}, 012502 (2009).

\bibitem{Bhatta} M.\ Bhattacharya {\it et al.}, \prc {\bf 77},
065503 (2008).

\bibitem{DW} D.\ H.\ Wilkinson, Ed., ``Isospin in Nuclear Physics'',
North Holland, 1969.
\end{thebibliography}
\end{document}